\documentclass[prd,reprint,nofootinbib]{revtex4-1}
\usepackage{caption,subcaption,multirow}
\usepackage{verbatim}
\usepackage{color}
\pdfoutput=1 

\usepackage[T1]{fontenc} 
\usepackage[normalem]{ulem}
\usepackage{color} 
\usepackage{amsfonts,amssymb,amsmath} 
\usepackage{mathtools} 
\usepackage{slashed,braket}       
\usepackage{bm}  
\usepackage{hyperref} 
\usepackage{parskip} 
\usepackage{empheq}
\usepackage{tensor} 
\usepackage{mathrsfs}
\usepackage[utf8]{inputenc}
\usepackage{bbold}

\usepackage{booktabs}
\definecolor{c1}{rgb}{0.09,0,0.61}
\definecolor{c2}{rgb}{0.7,0.1,0.7}
\definecolor{c3}{rgb}{0,0.5,0}
\definecolor{c4}{rgb}{0.7,0.3,0.1}

\DeclareMathAlphabet\mathbfcal{OMS}{cmsy}{b}{n}

\newcommand{\beq}{\begin{equation}}
\newcommand{\eeq}{\end{equation}}

\renewcommand{\subsubsection}[1]{\addtocounter{subsubsection}{1}
\par\nobreak
\medskip
\nobreak
\noindent{\it \thesubsubsection.  #1 }
\par\nobreak\medskip\nobreak}

\def\lpar#1#2#3#4{\rlap{\raise#3\hbox{$\hskip#4#1\left\{\mbox{\phantom{\rule[0mm]{0mm}{#2}}}\right.$}}}
\def\rpar#1#2#3#4{\rlap{\raise#3\hbox{$\hskip#4\left\}#1\mbox{\phantom{\rule[0mm]{0mm}{#2}}}\right.$}}}

\begin{document}
\title{Novel kinematics from a custodially protected diphoton resonance}
\author{Jack H. Collins}
\email{jhc296@cornell.edu}
\author{Csaba Cs\'aki}
\email{csaki@cornell.edu}
\author{Jeff A. Dror}
\email{ajd268@cornell.edu}
\author{Salvator Lombardo}
\email{sdl88@cornell.edu}
\affiliation{Department of Physics, LEPP, Cornell University, Ithaca, NY 14853, USA}

\begin{abstract}
We study a simple, well-motivated model based on a custodial symmetry which describes the tree-level production of a 750 GeV diphoton resonance from a decay of a singly produced vector-like quark. The model has several novel features. The identification of the resonance as an SU(2)$_R$ triplet provides a symmetry explanation for suppression of its decays to $hh$, $WW$, and $gg$. Moreover, the ratio of the 13 TeV to 8 TeV cross sections can be larger than single production of a 750 GeV resonance, reaching ratios of up to 7 for TeV scale vector-like quark masses. This eliminates any tension between the results from Run I and Run II diphoton searches. Lastly, we study the kinematics of our signal and conclude that the new production mechanism is consistent with available experimental distributions in large regions of parameter space but, depending on the mass of the new vector-like quarks, can be differentiated from the background with more statistics.
\end{abstract}
\maketitle

\section{Introduction\label{sec:Intro}}
The recent observation of an excess in the diphoton channel around 750 GeV invariant mass by ATLAS and CMS at $\sqrt{s} = 13$ TeV~\cite{ATLAS-CONF-2015-081,CMS-PAS-EXO-15-004} has generated much interest in models with a heavy scalar resonance, $\phi$, that decays to two photons. Most explanations proposed so far are considering loop induced resonance production, typically via heavy vector-like quarks (VLQ) charged under the Standard Model (SM). Otherwise, tree-level decays to SM particles would naturally dominate the branching ratio of $\phi$, either leading to a diphoton rate too small to explain the excess or a production rate of two SM particles with large invariant mass that is excluded by existing measurements. 

In this paper, we propose a novel tree-level production mechanism where $ \phi $ arises from the decay of a VLQ. The VLQs can be singly produced due to their mixing with the SM quarks, while the resonance is protected by the SU(2)$_L \times $SU(2)$ _R $ custodial symmetry.
In order to have significant mixing between the VLQs and the light quarks without modifying the $Z q \bar{q}$ couplings predicted by the SM, we introduce VLQs in a bidoublet representation of the custodial symmetry, while the resonance $\phi$ is part of a triplet under SU(2)$_R$. 
 The model has several advantages and new features:

\begin{itemize}
\item It is one of the few viable examples of tree-level production consistent with the excess signal rate, existing experimental constraints, and the kinematic distributions of the diphoton background events.
\item The ratio of production rates between 13 and 8 TeV is different than gluon or quark fusion. Depending on the model the ratio can be as large as about 7 (vs. 4.7 for gluon fusion), eliminating the tension with the 8 TeV diphoton searches.
\item The custodial symmetry protects the resonance from the leading one-loop decays to $hh$, $WW$, and $ g g $, while allowing decays to $\gamma\gamma$. The suppression of the $hh$ decay is particularly significant since in most models this coupling will arise at tree-level, making it difficult to reconcile with the expectantly large diphoton branching ratio and unobserved $ h h $ decays. The $\gamma\gamma$ (as well as $ ZZ $ and $ Z \gamma $) decay width is nonvanishing due to the explicit breaking of the custodial symmetry from gauging the U(1)$_Y$ subgroup of SU(2)$_R$.
\item The custodial symmetry also forbids one-loop gluon fusion production, explaining the dominance of the tree-level production via a decay of a VLQ.  
\end{itemize}

The scenario where $\phi$ is produced primarily from the decay of singly-produced VLQs has not been considered in the diphoton excess literature, although some authors have considered production of $\phi$ through a cascade decay of a heavier parent particle (e.g.,~\cite{Franceschini:2015kwy,Knapen:2015dap,Han:2016bus,Huang:2015evq,Liu:2015yec,Berlin:2016hqw,Bernon:2016dow,DeRomeri:2016xpb}). Furthermore, several authors have pointed out the potential to measure VLQ top or bottom partners decaying to $\phi$~\cite{Agrawal:2015dbf,Kobakhidze:2015ldh,Dev:2015vjd}. However, to explain the bulk of the excess signal through a decay of pair-produced top or bottom partners would require couplings at their perturbative limits to achieve large enough $\gamma\gamma$ rate and to explain why the VLQ decay process dominates over gluon fusion~\cite{Han:2016bus}. Moreover, pair production of two VLQs per event would give several hard jets in the event in addition to the diphoton, which is inconsistent with the kinematic distributions of events in the excess region. The tree-level production mechanism presented in this paper avoids these problems. Interestingly, it has been pointed out in~\cite{Delaunay:2013pwa} that a top partner need not be a mass eigenstate but rather could be a mixture of top and charm-like mass eigenstates raising the possibility that the vector-like quarks which mix with light quarks could also play a role in solving the little hierarchy problem.

This paper is organized as follows. In Sec.~\ref{sec:model}, we introduce a motivated model consistent with low energy flavor constraints in which a bidoublet of vector-like quarks with $m_{V} \gtrsim m_{\phi}$ mix with the light-SM quarks. In Sec.~\ref{sec:pheno} we discuss the production and decays of the new particles and find that the model can easily accommodate the current excess in the diphoton data without tension from existing searches, both from the $ 8 \text{ TeV} $ searches sensitive to the 750 GeV resonance as well as searches sensitive to VLQs. Furthermore, we compare the kinematic distributions of our signal and the diphoton background, finding that, depending on the splitting between the VLQ and the scalar, the distributions can be challenging to distinguish without additional data.

\section{Custodial symmetry and light quark mixing}\label{sec:model}
\begin{figure} 
  \begin{center} 
\includegraphics[width=6.5cm]{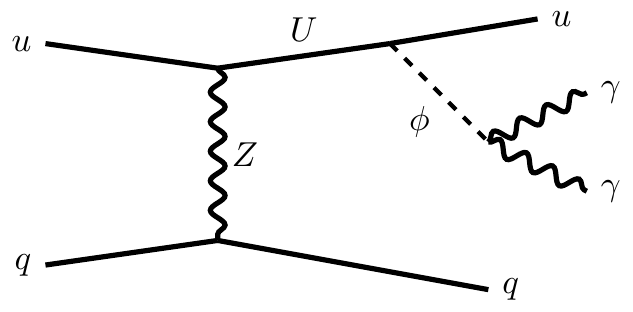} 
\end{center}
\caption{The dominant production of the diphoton excess from a decaying VLQ ($U$). In addition to the resonance, there are two additional jets. The $ p _T $ of the jets (and hence the visibility of the signal) is strongly dependent on the mass of the VLQ.}
\label{fig:prod}
\end{figure}

In this section, we present a model in which the resonance is a decay product of an electroweak produced VLQ from the dominant $ t $-channel process shown in figure~\ref{fig:prod}. Single production of VLQs requires a large mixing angle between the light quarks and VLQs. One would naively expect such mixing to yield large corrections to the $Z q \bar{q}$ couplings, which are strongly constrained by electroweak precision observables. However, we can protect the $Z$ couplings via the custodial symmetry by using a bidoublet representation $V$ for the VLQs~\cite{Agashe:2006at}. In addition, we introduce the SU(2)$_R$ triplet, $\Phi$, whose neutral component (which we denote by $ \phi $) will play the role of the 750 GeV resonance. This model is an example of a custodially symmetric model commonly considered in composite Higgs and extra dimensional models~\cite{Atre:2008iu,Atre:2011ae,Delaunay:2013pwa,Redi:2011zi,Flacke:2013fya,Redi:2013eaa,Vignaroli:2012nf,Vignaroli:2012sf,Atre:2013ap}.
\subsection{Field content and mixing}
We organize the fields into irreducible representations of SU(3)$ _C \times $SU(2)$ _L \times $SU(2)$ _R \times $U(1)$ _X $ in Table~\ref{tab:fields}, where the SU(2)$_R$ is a global symmetry and the fermions are left-handed Weyl spinors. Hypercharge is embedded in SU(2)$_R \times $U(1)$_X$ as $Y = T^3_R + X$. The SM quarks are taken to be singlets under SU(2)$_R$ and we represent the Higgs doublet as a bifundamental, $\mathcal{H} = (\epsilon H^*, H)$. 


Four VLQs form the bidoublet,
\begin{align}
& V \equiv 
\left( \begin{array}{c c} 
U_1 & X   \\  
D & U_2   
\end{array} \right), \ 
\overline{V} \equiv 
\left( \begin{array}{c c} 
\overline{U}_1 & \overline{D}   \\  
\overline{X} & \overline{U}_2   
\end{array} \right) \,.
\end{align}
The charges of the new quarks are $Q_{U_1} = Q_{U_2} = 2/3$, $Q_D = -1/3$, and $Q_X = 5/3$. 

\begin{table}
\center
\begin{tabular}{lcccc}
\toprule
\text{Field} & $\text{SU}(3) _C$ & $\text{SU}(2) _L$ & $\text{SU}(2) _R$ & $\text{U}(1) _X$ \\ \hline 
$\Phi$ & ${\mathbf{1}}$ & ${\mathbf{1}}$ & ${\mathbf{3}}$  &  0 \\  
$V$ & ${\mathbf{3}}$ & ${\mathbf{2}}$ & ${\mathbf{2}}$ & $+ 2/3$  \\
$\overline{V}$  & $\bar{{\mathbf{3}}}$ & ${\mathbf{2}}$ & ${\mathbf{2}}$ & $- 2/3$  \\ 
\hline \vspace{-0.3cm} \\ 
${\cal H}$  & ${\mathbf{1}}$  & ${\mathbf{2}}$ & ${\mathbf{2}}$ & 0 \\ 
$Q$ & ${\mathbf{3}}$ & ${\mathbf{2}}$ & ${\mathbf{1}}$ & $+ 1/6$ \\ 
$\overline{u}$   & $\bar{ {\mathbf{3}}}$  & ${\mathbf{1}}$ & ${\mathbf{1}}$ & $- 2/3$\\
\bottomrule
\end{tabular}
\caption{The representations of relevant fields. The new vector-like quarks are in a single bidoublet $V$, and the 750 GeV resonance is the neutral component of $\Phi$. The Higgs and the light quarks have the usual SM assignments. All fermionic fields are left-handed Weyl spinors.}
\label{tab:fields}
\end{table}

The Lagrangian for the VLQs (not including the terms coupling to the scalar $\Phi$) is 
\begin{equation} 
{\cal L} _{ VLQ} = m _V  \text{Tr} \left[ \overline{V} V  \right]   - \lambda _V \text{Tr}  \left[ {\cal H} ^\dagger  V  \right]    \overline{u} ^{(0)} + h.c.\,,
\end{equation} 
where the $(0)$ superscript denotes the quark fields in the SM mass basis of the up-type sector (the basis of diagonal SM Yukawa couplings). We assume the VLQs only mix with a single generation of right-handed up-type quark, however in appendix~\ref{app:down} we also consider a different U(1)$_X$ charge for $V$ and the case of mixing with one generation of right-handed down-type quark. Note that we have made an important assumption regarding alignment: the bidoublet $V$ couples to the up-type quark in the mass basis of the SM. This assumption is to avoid low energy flavor constraints from flavor changing neutral currents but is not a crucial ingredient for the collider phenomenology that follows. 

The down-type quark masses are unaffected by the new VLQs. One flavor of the up-type quarks can mix significantly with the new VLQs through the off-diagonal mass matrix:
\begin{equation} 
\left( \begin{array}{ccc}u ^{(0)}  & U_1  & U_2 \end{array} \right) \left( \begin{array}{ccc} 
\lambda _u v / \sqrt{2}   & 0 & 0 \\  
- \lambda _V v / \sqrt{2}   & m _V  & 0 \\  
- \lambda _V v / \sqrt{2}   & 0 & m _V   
\end{array} \right) \left( \begin{array}{c} 
\overline{u}^{(0)}   \\  
\overline{U}_1  \\  
\overline{U}_2  
\end{array} \right) + h.c.
\label{eq:mixing}
\end{equation} 
where $ \lambda  _u $ is the Yukawa of the up-type quark. From here on, we assume the mixing is with the up or charm quark and neglect the up-type quark mass. The mass eigenstates ($u$, $U$, and $\tilde{U}$) are related to the gauge eigenstates by:

\begin{align} 
\left( \begin{array}{c} 
\overline{u} ^{(0)}  \\  
\overline{U}_1  \\  
\overline{U}_2 
\end{array} \right) & = \left( \begin{array}{ccc} 
1 & 0 & 0 \\  
0 & 1 / \sqrt{2}  & -1/ \sqrt{2}  \\  
0 & 1 /\sqrt{2}  & 1 /\sqrt{2}   
\end{array} \right) \left( \begin{array}{ccc} 
c _\theta  & -  s _\theta  & 0 \\  
s _\theta  & c _\theta  & 0 \\  
0 & 0 & 1  
\end{array} \right) \left( \begin{array}{c} 
\overline{u}   \\  
\overline{\tilde{U}} \\  
\overline{U} 
\end{array} \right) \label{eq:Rmix}  \\ 
 \left( \begin{array}{c} 
u ^{(0)}  \\  
U_1  \\  
U_2  
\end{array} \right) & = \left( \begin{array}{ccc} 
1 & 0 & 0 \\  
0 & 1 / \sqrt{2}  & - 1/ \sqrt{2}  \\  
0 &  1 /\sqrt{2}  & 1 /\sqrt{2}   
\end{array} \right) \left( \begin{array}{c} 
u  \\  
\tilde{U}  \\  
U  
\end{array} \right) \notag
\end{align}  
where $ s _\theta \equiv \sin \theta $ and $ c _\theta \equiv \cos \theta $ with
\begin{equation} 
s _\theta = \frac{  \lambda _V v }{ m _{ U ^+ }}\,,
\end{equation} 
where $ v \simeq 246 \text{ GeV}$ is the vacuum expectation value of the Higgs, and the masses of the VLQs are $m _{ U ^- } = m _{ D } = m _{ \chi } = m _V $ and $m _{ U ^+ } = \sqrt{ m _V ^2 + \lambda _V ^2 v ^2 } $. The down-sector does not experience any mixing, i.e. $X$ and $D$ are mass eigenstates. 

The mixing, parameterized by $ s _\theta$, leads to couplings between a generation of SM quarks, SM gauge bosons, and the VLQs (derived in more detail in appendix~\ref{app:couplings}):
\begin{equation}
{\cal L} _{EW} = -\frac{e s_\theta }{2 c_w s_w} Z_\mu \overline{ u} ^\dagger \bar{\sigma} ^\mu \overline{ U} - \frac{g s_\theta}{2  } W^{-}_\mu( \overline{u} ^\dagger \bar{\sigma} ^\mu \overline{D} + \overline{X} ^\dagger \bar{\sigma} ^\mu \overline{u} ) + h.c.\,,
\end{equation}
where $ s _w $ ($ c _w $) is the sine (cosine) of the Weinberg angle and $ g $ ($ e $) is the SU(2)$_L$ (QED) coupling constant. If the mixing angle is sufficiently large, these couplings can result in electroweak production of single VLQs ($U$, $D$, or $X$) which can dominate over the VLQ pair production cross section. Notice that only $U$, which will be responsible for the production of the diphoton resonance, couples to an up-type quark and the $Z$. $\tilde{U}$ is not produced by electroweak interactions in this model. This is because only the linear combination $\overline{ U}_1 + \overline{ U}_2$ mixes with the up-type quark, which is a necessary feature for the protection of the $Z q\bar{q}$ coupling in this model.

\subsection{Consequences of a custodial triplet}

The 750 GeV diphoton resonance $\phi$ is embedded in an SU(2)$_R$ triplet scalar $\Phi$ as follows.
\begin{equation} 
\Phi = \left(\  \begin{array}{cc} 
\phi / \sqrt{2} & \phi _+  \\  
\phi _-  & - \phi / \sqrt{2} 
\end{array} \right)
\end{equation}
This allows for a coupling of $\Phi$ to the VLQs of the form
 \begin{align} 
{\cal L} _{\Phi}&  = \sqrt{2} y_\phi \text{Tr} \left[ \overline{ V}  \Phi  V \right]  + h.c. \\ 
 & =  y_\phi \phi   \left( U_1 \overline{U}_1  - U_2 \overline{U}_2  + X \overline{X} - D\overline{D}  \right)   + h.c. + ... \label{equ:cancellation}\\ 
& = y _\phi \phi \left( -s _\theta U \bar{u} - \tilde{U} \overline{ U} - c _\theta U \overline{ \tilde{U} } + X \overline{ X} - D \overline{D}\right)  \notag \\ 
& \hspace{5cm} + h.c. + ... \label{eq:USu}
\end{align} 
where the ellipses refer to terms involving the charged components of $\Phi$. The relative minus sign between the $  U _1 \overline{ U } _1 $ and $ U _2  \overline{ U } _2 $ terms gives rise to the coupling of $ \phi $ to $ U $ and the SM up quark (as opposed to a coupling to $ \tilde{U} $)  which is responsible for the production of the resonance.

These interactions will generate couplings of $\phi$ to SM dibosons, such as $gg$, $WW$, $hh$ etc. via triangle diagrams with the VLQs. However, in the limit of exact custodial symmetry, these amplitudes are forbidden. For example, the operator $\Phi \, G^{\mu \nu, A} G_{\mu \nu}^A$ has no custodially invariant contraction because the gluon field strength tensor $G$ is a custodial singlet. Furthermore, the Higgs coupling to the scalar vanishes since (using ${\cal H H ^\dagger } \propto \mathbb{1} $), 
\begin{equation} 
\text{Tr} \left[ {\cal H} ^\dagger  \Phi {\cal H}  \right]  \propto \text{Tr}\, \Phi  = 0  \,.
\end{equation} 
In practice, the vanishing of the amplitudes is a consequence of cancellations of the contributions due to different VLQs running in the loop, which contain important relative minus signs as a consequence of the custodial symmetry.

The loop amplitudes for $\phi$ therefore require the insertion of SU(2)$_R$ violating interactions. The largest such couplings in the SM are the third generation Yukawas, however in the flavor alignment limit these couplings will not directly affect the diphoton resonance sector since we assume mixing is not occurring with the third generation up-type quark. The dominant source of custodial symmetry breaking in this sector will therefore be the embedding of the hypercharge gauge group within the $T_3^R$ generator of SU(2)$_R$, and so it is to be expected that the leading loop amplitude will be that coupling $\phi$ to hypercharge gauge bosons, $\phi \, B^{\mu \nu} B_{\mu \nu}$. Indeed, the one-loop contributions to this operator do not cancel among the VLQs due to their differing hypercharges. The other loop amplitudes will be generated at higher order and are suppressed by an additional factor $\sim \alpha / 4 \pi c_w^2$ compared to their naive sizes. These two-loop contributions can induce a mixing angle between $ \phi $ and the Higgs of order $ \sim ( v  / 16 \pi ^2 m_\phi  )  \,( \alpha  / 4 \pi c _w ^2  ) $ (where $ m _\phi $ denotes the mass of the resonance), however this is much too small to induce sizable decays to $ t \bar{t} $. Direct couplings of $ \phi  $ to the up-type quarks can also arise at two-loops but is suppressed by a Yukawa coupling, making it negligible. We verify the effects of custodial symmetry breaking explicitly in~\ref{app:custodial_breaking}. This custodial protection mechanism generates a natural hierarchy between the decays of the resonance to diphotons and its decays to $gg$, $WW$, $hh$, and also suppresses the gluon fusion production of $\phi$.

We now briefly compare this to a scenario in which the diphoton resonance is assumed to be a custodial singlet $S$ with couplings  
 \begin{align} 
{\cal L} _{S}& = y_S S \, \text{Tr} \left[  \overline{V} V \right]  + h.c. \\ 
& =   y_S S \left( U_1 \overline{U}_1  + U_2 \overline{U}_2 + D \overline{D} + X \overline{X} \right) + h.c. \label{equ:nocancellation} \\ 
& =  y_S S\left(  s _\theta \tilde{U}  \overline{u} + U \overline{U}  + c  _\theta \tilde{U} \overline{\tilde{U}} + D \overline{D} +  X \overline{X} \right) + h.c.
\end{align}
In this case, the only $S$-quark-VLQ coupling involves $\tilde{U}$, which does not couple to SM gauge bosons and therefore cannot be produced via VLQ single production. $\tilde{U}$ is pair-produced and can decay $\tilde{U} \rightarrow S u$, however the rate for pair production is subdominant to electroweak production of $U$ and insufficient to explain the excess.
Furthermore, this singlet does not exhibit custodial protection, which is a consequence of the couplings in eq.~\ref{equ:nocancellation} adding constructively rather than destructively.

\section{Diphoton cross section}
\label{sec:pheno}
Above we presented a model in which $\phi$ can be produced as a decay product of a singly produced VLQ (we will assume $ m _V > m _\phi $ throughout). The dominant production mechanism for the diphoton resonance is depicted in figure~\ref{fig:prod}.\footnote{Secondary production modes from $ U Z $ production, pair production of VLQs, and direct $ \phi Z $ production mediated by a VLQ make up 10-30\% of inclusive diphoton cross section. We explore the size of different contributions in Sec.~\ref{sec:additionalfeatures}, but since the size of the subdominant modes is highly dependent on the detailed parameters of the model, we only include the dominant production when studying the inclusive diphoton rate and kinematics.} In this section, we demonstrate that the $\gamma\gamma$ rate is sufficient to explain the diphoton excess while avoiding constraints from existing VLQ searches and electroweak precision tests. We consider two variations of the model, one where the VLQs mix with the up quark and another with the VLQs mixing with the charm quark. 

Since the $ \gamma \gamma $ final state arises from a decay chain, the inclusive cross section into $ \gamma \gamma $ is given in the narrow width approximation by
\begin{equation} 
\sigma  _{ \gamma \gamma }= \sigma ( p p \rightarrow U \bar{u} , \bar{U}  u ) \times \text{Br} ( U \rightarrow \phi  u ) \times \text{Br} ( \phi \rightarrow \gamma \gamma ) \label{eq:NWE}.
\end{equation} 
Each of these contributions has different dependence on the relevant parameters of the model, $ y _\phi $ and $ s_\theta $. The production cross section of the VLQs, $ \sigma ( p p \rightarrow U \bar{u} , \bar{U} u )$, is proportional to $  s _\theta ^2 $ but is independent of $ y _\phi $.  
\subsection{Branching ratios}
The complete formulae for the branching ratios of $\phi$ and $U$ are given in appendix~\ref{app:decays}. We summarize the results here. $ U $ has two decay channels, $ U \rightarrow Z j $ and $ U \rightarrow \phi j $ with the dominant decay being $ Z j $. This results in the branching ratio of $U \rightarrow u \phi$ ranging between 1-10\%, proportional to $ y _\phi ^2 $ and independent of the mixing angle. $ \phi $ has competing decays between a $ 3 $-body tree-level decay and loop-induced 2-body decays. The only tree-level decay of $ \phi $ is to $ Z u \bar{u} $ through an off-shell $ U $ with a rate is proportional to $ s _\theta ^4 y _\phi ^2  $, making it highly sensitive to the mixing angle. $ \phi $ has additional loop-induced decays into $ \gamma \gamma $, $ Z \gamma $, and $ ZZ $. These decays arise from gauging hypercharge resulting in the relative ratios,
\begin{equation} 
\Gamma _{ \gamma \gamma } \, : \, \Gamma _{ \gamma Z} \, : \, \Gamma _{ ZZ} = 1 \, : \, 2 \tan ^2 \theta _w \, : \, \tan ^4 \theta _w \,.
\end{equation} 
The loop-induced rates are largely independent of the mixing, proportional to $ y _\phi ^2 $.\footnote{We have checked that non-zero mixing has at most a 10\% effect on the loop-induced rates in the region of parameter space we are interested in. Furthermore, these effects will not have any bearing on the size of the $ \gamma \gamma $ rate and thus we ignore these effects in our analysis.}

\begin{table}
\center
\begin{tabular}{lccc}
\toprule
 & $ y _\phi  $& $ s _\theta $  \\ \midrule 
$u _R \text{ benchmark}$ &0.7 & 0.1  \\
$c _R \text{ benchmark}$ &2 & 0.3   \\
\bottomrule
\end{tabular}
\caption{Benchmark points for the up quark and charm quark mixing models. For the $ u _R $ model, there are more stringent constraints on the mixing angle though a larger production cross section.} 
\label{tab:BM}
\end{table}

\begin{figure*}[!t] 
  \begin{center} 
\includegraphics[width=7cm]{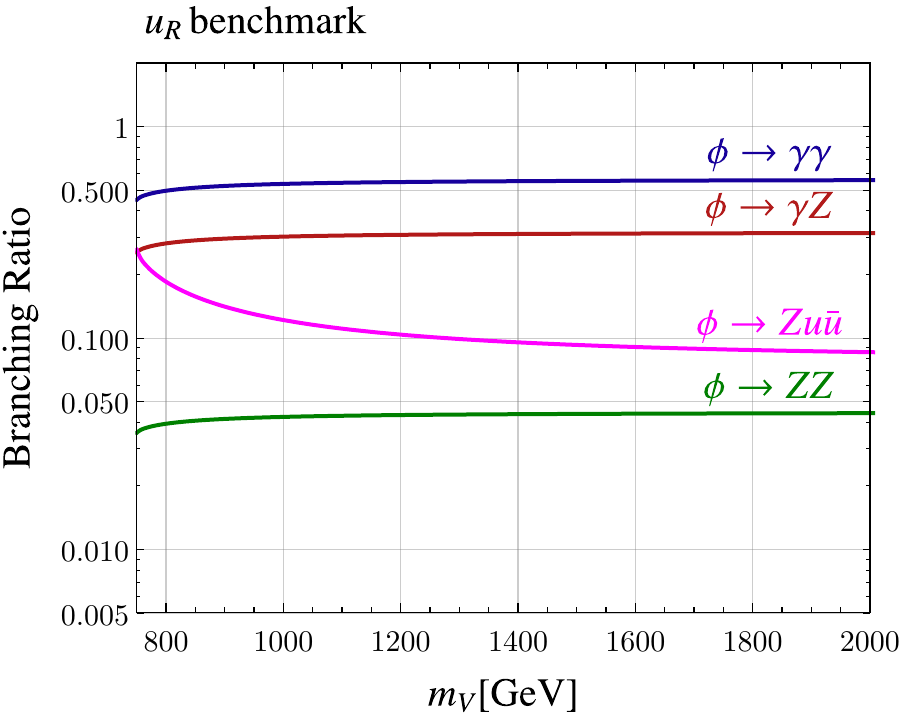} 
\includegraphics[width=7cm]{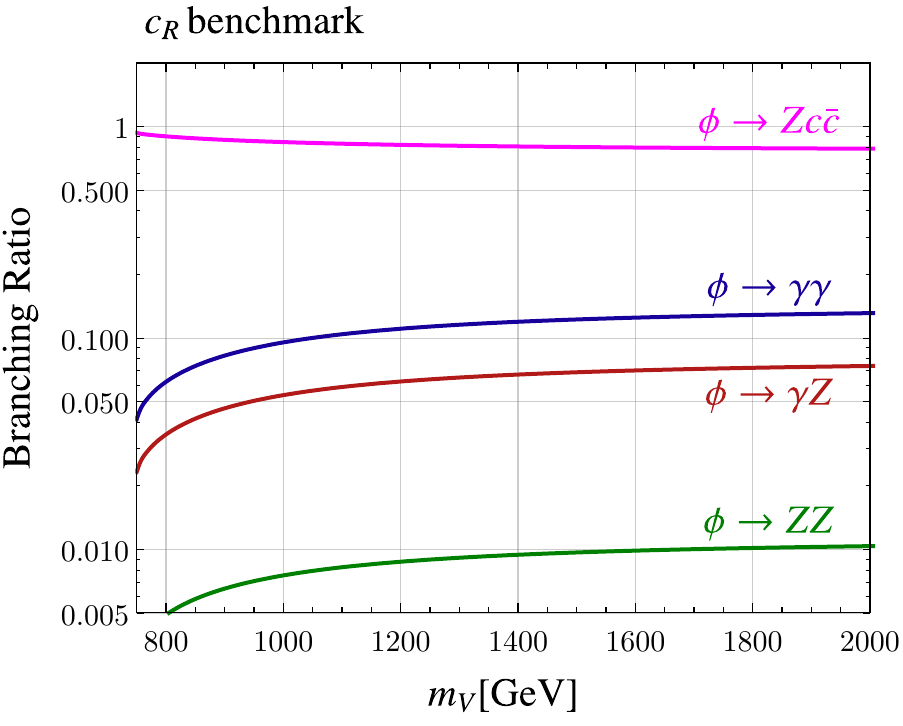} 
\end{center}
\caption{The different branching ratios for $ \phi $. The loop-induced decays to $ \gamma \gamma $, $ \gamma Z $, and $ ZZ $ always compete with the $ 3 $-body decay. At large mixing, the $ 3 $-body decay is the preferred decay mode, however for small-mixing, the loop-induced decays (which are roughly independent of the mixing) dominate. \textbf{Left:} The branching ratios for the $ u _R $ benchmark point (small-mixing).  \textbf{Right:} The branching ratios for the $ c _R $ benchmark (large-mixing).}
\label{fig:phibranching}
\end{figure*}

The leading branching ratios of $\phi$ are shown in figure~\ref{fig:phibranching} for a benchmark point relevant in the case of up-mixing with $s_\theta = 0.1$ 
(left) and charm-mixing with $s_\theta = 0.3$ (right). A couple comments on the choice of benchmark points (displayed in Table~\ref{tab:BM}) are in order. First, regarding the size of the mixing angle, we have provided constraints on the allowed mixing angle for electroweak produced VLQs in appendix~\ref{app:constraints}, obtained by reinterpreting the constraints from direct LHC searches on light quark composite partner models~\cite{Delaunay:2013pwa}. For the up-quark mixing, the mixing angle is experimentally constrained to be $ s _\theta \lesssim 0.12 $, while for charm mixing, the constraints are much weaker with $ s _\theta \lesssim 0.5 $. Larger mixing angles are allowed for the case of charm mixing since the electroweak production cross section of the VLQs are suppressed by the charm parton distribution function. 
\subsection{The inclusive cross section}
Depending on the size of mixing, the inclusive cross section for diphoton production scales differently with the mixing angle. There are two distinct regimes, large and small mixing angles. If the mixing angle is large, then the dominant decay of $ \phi $ is through the $ 3 $-body decay. In this case the branching ratio into diphotons is $ \propto 1/s _\theta ^4 $ giving an inclusive cross section
\begin{equation} 
\sigma _{ \gamma \gamma } \propto \frac{ y  _\phi ^2 }{ s _\theta ^2}  \quad \left( \text{large mixing} \right) .
\end{equation} 
For small mixing, the $3$-body $ \phi $ decay is heavily suppressed making the diphoton rate the dominant mode, i.e. $ \text{Br} ( \phi \rightarrow \gamma \gamma )  \approx 1$, independent of $ y _\phi $ or $ s _\theta $. In this case, the inclusive cross section scales as
\begin{equation} 
\sigma _{ \gamma \gamma } \propto y _\phi ^2 s _\theta ^2 \quad \left( \text{small mixing} \right). 
\end{equation} 
The transition between the two regimes occurs around $ s _\theta  \sim 0.2 $, and this is the point where the cross section is maximized. Due to the constraints, the up mixing case is always in the small mixing scenario while the charm can be either the large or small mixing regimes. For our chosen benchmark point, the charm scenario corresponds to large mixing.

\begin{figure*} 
  \begin{center} 
\includegraphics[width=7cm]{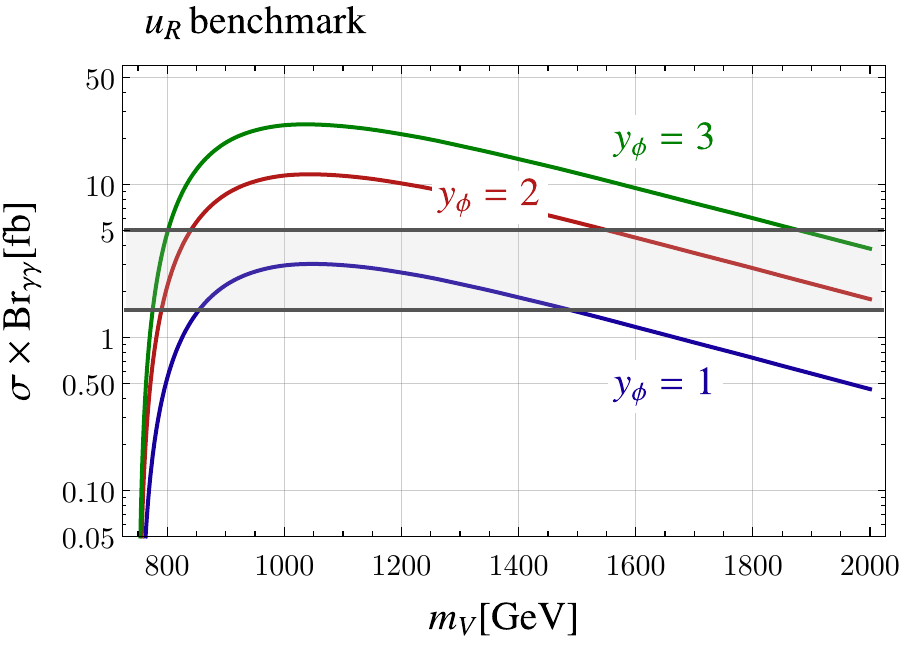}
\includegraphics[width=7cm]{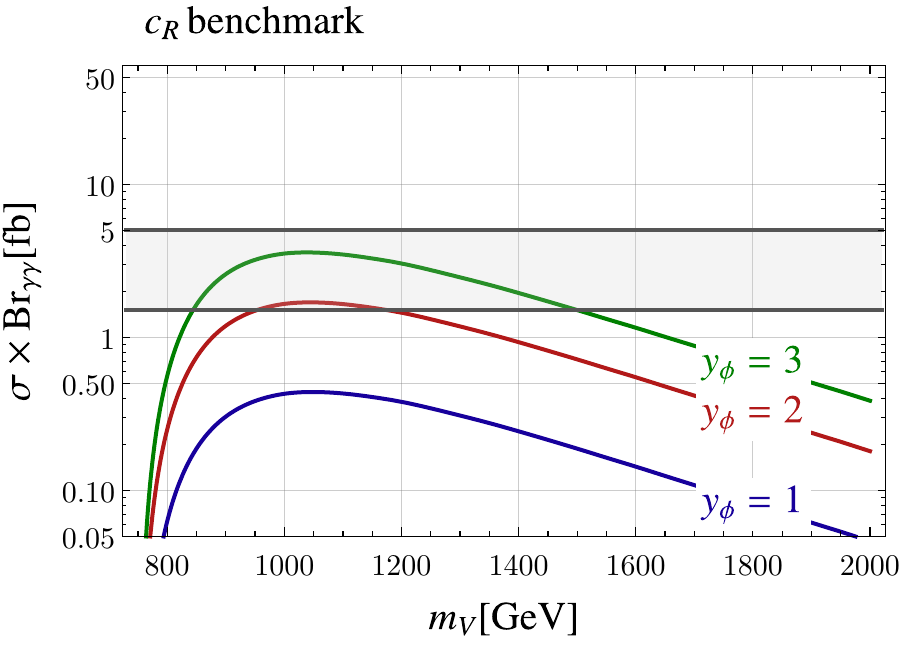}
\end{center}
\caption{The inclusive cross section into $ \gamma \gamma $ as a function of VLQ mass and varying the values of $ y _\phi  $. In gray we show the rough cross section necessary to explain the excess with a narrow width ($ 1.5  - 5 ~\text{fb} $)~\cite{Kamenik:2016tuv}. \textbf{Left:} The cross section for the $ u _R $ benchmark point. \textbf{Right:} The cross section for the $ c _R $ benchmark point. }
\label{fig:XSBranching}
\end{figure*}

To reproduce the excess, we simulate the production of $\phi$ at leading order using a custom {\scshape FeynRules} model~\cite{Alloul:2013bka} with {\scshape MadGraph5}~\cite{Alwall:2014hca}. To roughly estimate the size of next-to-leading order (NLO) effects, we compute the production cross section with an additional jet, finding that it makes up about 50\% of the leading order cross section. This suggests an NLO $K$-factor of about 1.5, and we use this correction throughout.    
We compute the diphoton rate using equation~\ref{eq:NWE} and the branching ratios given in appendix~\ref{app:decays}. Figure~\ref{fig:XSBranching} shows the inclusive diphoton production cross section for different VLQ masses and values of $ y _\phi $ for the two benchmark points. 

We see in the left of figure~\ref{fig:XSBranching} that we need $ y _\phi  \sim  0.7 $ to get enough cross section to explain the excess in the up quark variation of the model. Larger Yukawa couplings are required in the charm-mixing benchmark point, requiring $y_\phi \sim 2$ to achieve the minimum cross section needed to explain the excess. 

As with many models explaining the diphoton excess, such Yukawa couplings can lead to non-perturbativity of the model before the GUT scale. The up-mixing benchmark becomes non-perturbative at around 100 TeV, although there is some parameter space where the coupling remains perturbative beyond the GUT scale. The charm-mixing model is more problematic given the Yukawa coupling runs to its perturbative limit at a few TeV, putting into question the validity of our analysis. However, this problem can be easily overcome by adding additional flavors of bidoublets (these may or may not mix with the SM quarks) which feed into the running, but also boost the diphoton decay rate as the square of the number of flavors allowing for much smaller couplings, and a much higher scale of strong coupling.

\begin{figure} 
  \begin{center} 
\includegraphics[height=8cm]{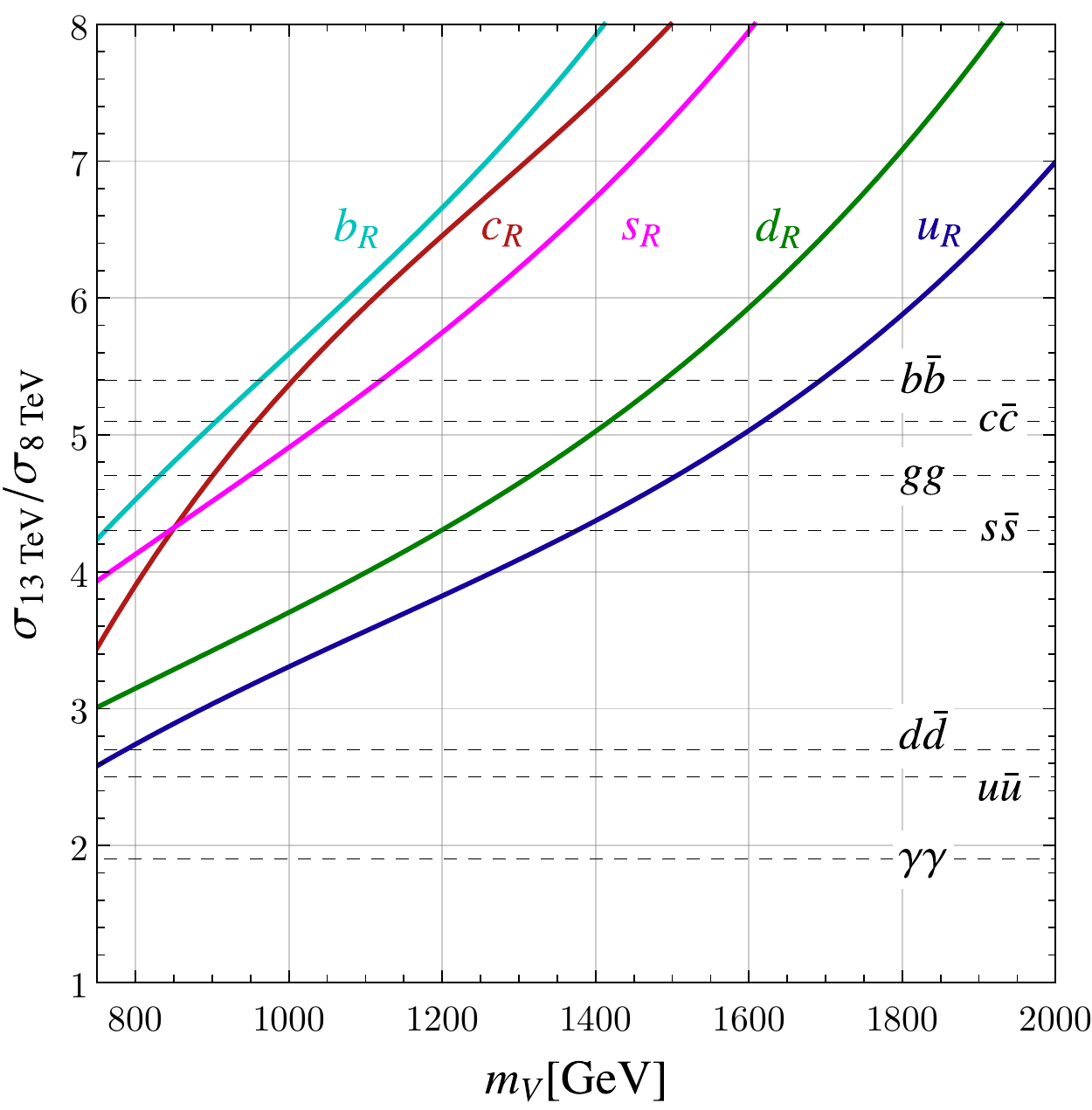} 
\end{center}
\caption{The ratio of $\sqrt{s} = 13 $ to $8$ TeV cross section of the up-mixing signal (blue) and charm-mixing signal (red). We have included the scaling properties of the down-type version of this model where the down (green), strange (pink), or bottom (light blue) quark mix with VLQs. The scaling of other proposed production processes are shown as dashed lines and were taken from~\cite{Franceschini:2015kwy}.}
\label{fig:Scaling}
\end{figure}

\subsection{Eliminating tension with $ 8 $ TeV data}
One of the puzzling features of the 13 TeV diphoton excess is its seemingly large cross section compared to cross section limits from 8 TeV searches. A 750 GeV resonance is not ruled out by Run I searches but, depending on the production mechanism and width, may be in tension with Run I limits~\cite{Falkowski:2015swt,Kamenik:2016tuv}. Thus to reproduce the excess, it is important to have sufficiently large scaling, $ r $, defined as the ratio of the cross section at 13 TeV to that at 8 TeV. We compute the scaling for our model as a function of the VLQ mass (the scaling is independent of the couplings) and assuming the $ K $-factor is constant from $ 8 $ to $ 13 $ TeV. The results are shown in figure~\ref{fig:Scaling} alongside the scaling of other proposed models, including gluon fusion, $q \bar{q}$ production~\cite{Gao:2015igz}, and photon fusion~\cite{Csaki:2015vek,Fichet:2015vvy,Csaki:2016raa}. The up-mixing model inherits the scaling from the $u \bar{u}$ production at $m_{V} \sim 750$ GeV but grows with the mass of the VLQ due to the higher center-of-mass energy. For heavy VLQ masses near 1500 GeV, the scaling is comparable to gluon fusion. The charm variation of the model, however, scales much better due to the parton distribution function of the charm in the initial state. For VLQ masses nearly degenerate with $m_\phi$ ($800$ GeV $\lesssim m_{V} \lesssim 1000 $ GeV) for which the extra jet from the $U$ decay is relatively soft, the scaling is as large as for $b\bar{b}$ production. For larger masses, the charm-mixing scenario has $r \gtrsim 7$, but in this region of parameter space, the $\phi$ would be accompanied by a high-$p_T$ jet in the final state (we explore the plausibility of this scenario in Sec.~\ref{sec:kinematics}). We conclude that, depending on the mass of the VLQ, our signal can achieve larger cross section scaling from 8 to 13 TeV than any proposed model of single resonance production. 

For simplicity, we have only considered mixing with up-type quarks. It is possible to construct a similar model in which new VLQs mix with the down-type quarks. The terms in the Lagrangian responsible for production of a down-type variation of this model are presented in appendix~\ref{app:down}. We include the scaling properties of production from down, strange, and bottom quark mixing in figure~\ref{fig:Scaling}. The bottom quark mixing scenario has the largest ratio of 13 to 8 TeV cross section, while the down and strange quark scenarios interpolate between the scaling of the up and charm scenarios. 
\section{Kinematics\label{sec:kinematics}}
\subsection{Comparing with ATLAS}
\begin{figure*} 
  \begin{center} 
\includegraphics[height=4cm]{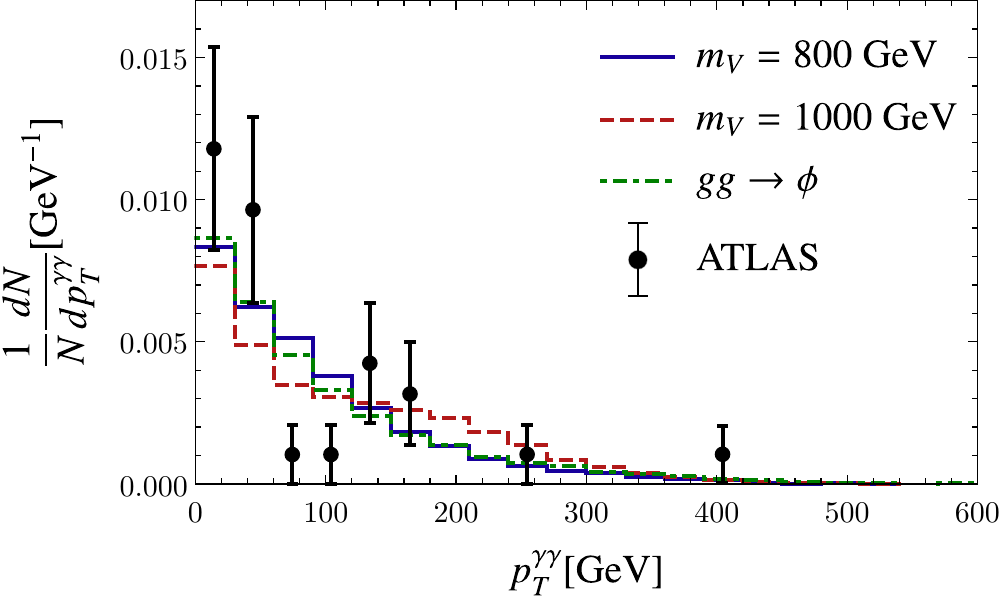} 
\includegraphics[height=4cm]{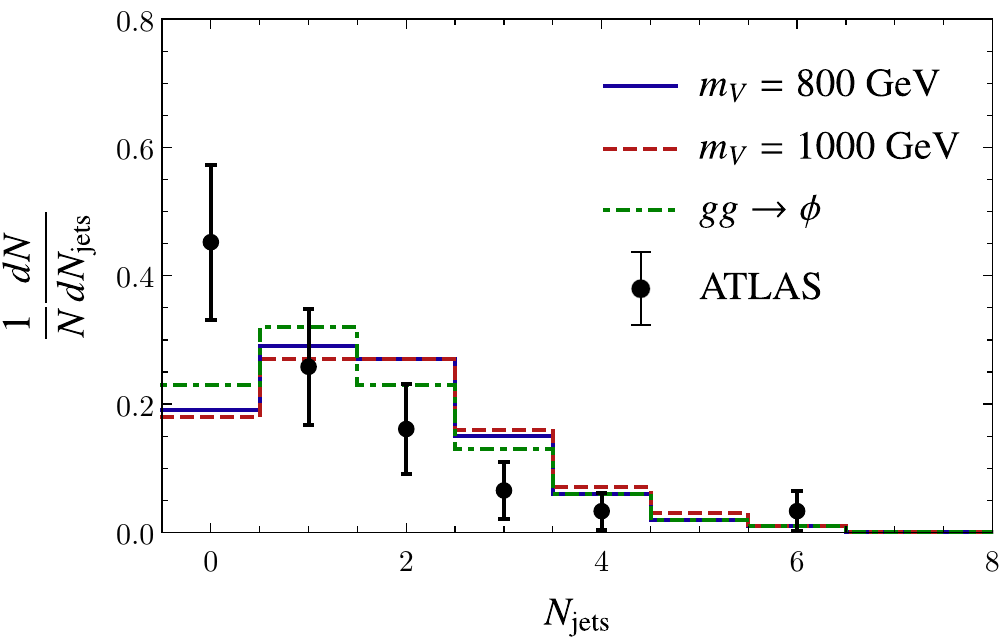} 
\end{center}
\caption{The kinematic distributions of the sum of the signal and background for vector-like quark mass of $ 800 $ (blue) and $ 1000 \text{ GeV} $ (red) compared to the distributions observed by ATLAS (black) with 3.2 fb$^{-1}$ of data. We also provide gluon fusion kinematics (green) for comparison. The $ N _{ \text{jets}} $ and $ p _T ^{ \gamma \gamma } $ distributions for the background and observed events are obtained from the slides presented by ATLAS~\cite{Moriond_ATLAS}.}
\label{fig:kinematics1}
\end{figure*}

\begin{figure*} 
  \begin{center} 
\includegraphics[height=4cm]{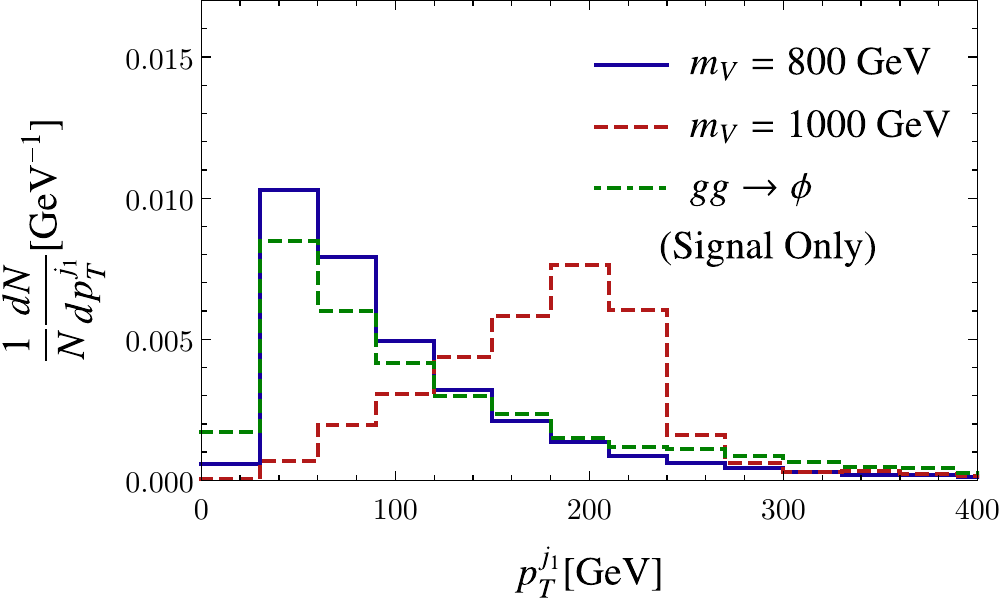}
\includegraphics[height=4cm]{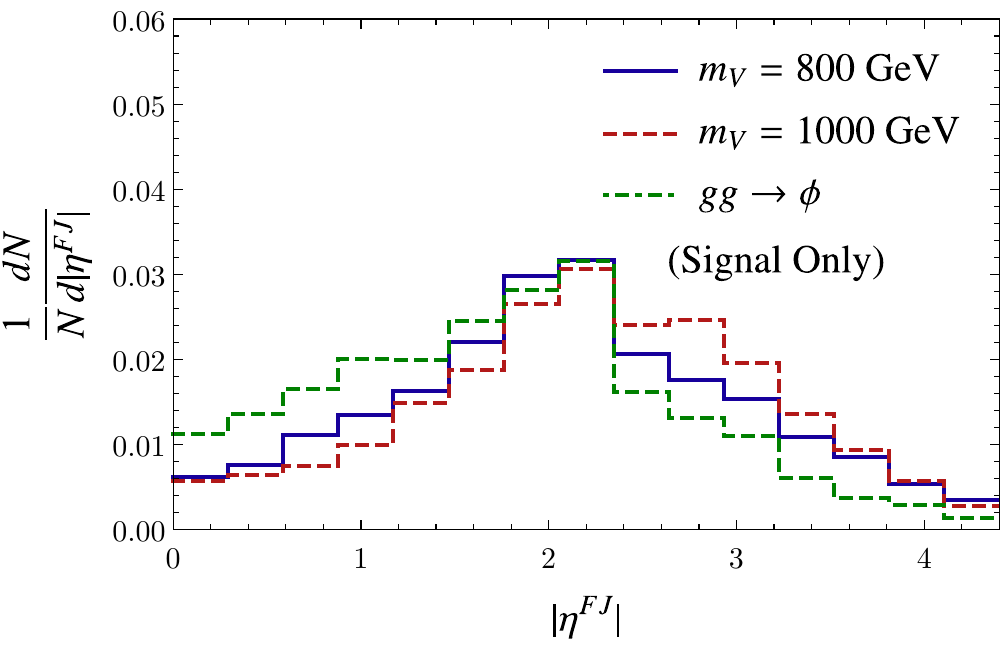}
\end{center}
\caption{The leading-jet $p_T$ and forward jet pseudorapidity distributions of the signal for vector-like quark mass of $ 800 $ (blue) and $ 1000 \text{ GeV} $ (red) along with a gluon fusion signal (green) for comparison. For VLQs almost degenerate with the resonance, the signal is difficult to differentiate from the QCD background or a resonance produced via gluon fusion since these events also contain soft, forward jets from initial state radiation.
}
\label{fig:kinematics2}
\end{figure*}

In addition to the diphoton resonance signature at $ m_{\gamma\gamma} \sim 750 \text{ GeV} $, our signal has two additional jets, with one of the jets typically in the forward direction.
ATLAS and CMS have remarked that events in the excess region are consistent with the background kinematics. Furthermore, ATLAS has recently provided kinematic distributions of the excess events~\cite{Moriond_ATLAS} for the number of jets, $ N _{ \text{jets}} $ (ATLAS defined a jet using $ p _T ^{ j } > 25 \text{ GeV}$ for $ \eta < 2.4 $ and $ p _T ^{ j } < 50 \text{ GeV} $ for $ \eta < 4.4 $), and the transverse momentum of the $ \gamma \gamma $ resonance, $ p _T ^{  \gamma \gamma } $. Furthermore, ATLAS provided estimates for the expected SM diphoton background from simulations. The distributions are provided for the region $ 700 \text{ GeV} <m _{ \gamma \gamma } < 840 \text{ GeV} $ and with the requirements on the leading and subleading photon energies $ E _T ^{ \gamma _1 } > 0.4 m _{ \gamma \gamma } $ and $ E _T ^{ \gamma _2 } > 0.3 m _{ \gamma \gamma } $. In this bin, ATLAS found a total of $ 34 $ events, about 10 of which are diphoton excess candidates. 

To compare the compatibility of the kinematics of the excess with our signal we simulate our signal using a combination of {\scshape MadGraph5}, {\scshape Pythia 8.2}~\cite{Sjostrand:2007gs}, and {\scshape Delphes 3}~\cite{deFavereau:2013fsa} making use of the {\scshape NNPDF2.3LO} parton distribution functions~\cite{Ball:2012cx}. To compare with the distributions observed by ATLAS, we perform a weighted sum of our signal and the background estimates provided by ATLAS
\begin{equation} 
N = r N _{ sig} + ( 1 - r )  N _{ bkg}.
\end{equation} 
This is done for each bin and we take $ r \approx 10/ 34 $. We also simulate gluon fusion at leading order and perform this procedure in order to compare our signal with the kinematics of single production. By comparing the distributions in figure~\ref{fig:kinematics1}, we conclude that although our signal has two extra jets in the final state, the distributions for $N _{ \text{jets}}$ and $ p _T ^{ \gamma \gamma } $ are consistent with the data provided by ATLAS (as is also the case for the gluon fusion signal). Furthermore, the mass of the VLQ has only a mild effect on these distributions since the number of jets from the hard process is independent of $m_V$. The signal does have distinctive features in other distributions, however, and we explore these features in the next section.

\subsection{Additional signatures}
\label{sec:additionalfeatures}
Our signal predicts observable jet signatures that can be used to discern this process from background events or from other resonance production mechanisms. In particular, we expect a forward jet as well as one central jet with higher $ p _T $, depending on $m_V$. The distributions for the $ p _T $ of the leading jet ($ p _T ^{ j _1 } $) and the absolute value of the pseudorapidity of the most forward jet ($ |\eta ^{ FJ}| $) for the signals with $m_V = 800$ and 1000 GeV are shown in figure~\ref{fig:kinematics2} along with the gluon fusion signal for comparison. Note that ATLAS and CMS did not provide the background or observed kinematic distributions for these observables, so we did not combine the background and the signal in these plots.

The distributions have some distinctive characteristics. Firstly, we see that for small splitting $ p _T ^{ j _1 } $ is peaked around zero since, at these splittings, the central jet, which will typically be the leading jet, has low $ p _T $. However, for larger splitting the distribution has a kinematic edge with the end-point at the splitting between the VLQ and the resonance. This is prototypical of a jet arising from a heavy particle decay to a second heavy particle. Interestingly, such a distribution can suggest the mass of the VLQ. 
The forward jet in the event is most easily probed using its pseudorapidity. The distribution has a dip at $ \eta \approx 2.4 $ as a consequence of the cuts used by ATLAS for the jet definition (see above). As expected, the signal has a jet with large $ \eta $, however this is also true of the dominant background, $ \gamma \gamma $. In this background, jets are emitted from the initial state and hence tend to be in the forward region. This can result in the feature being well hidden inside the SM background of the searches.

\begin{figure*} 
  \includegraphics[width=7cm]{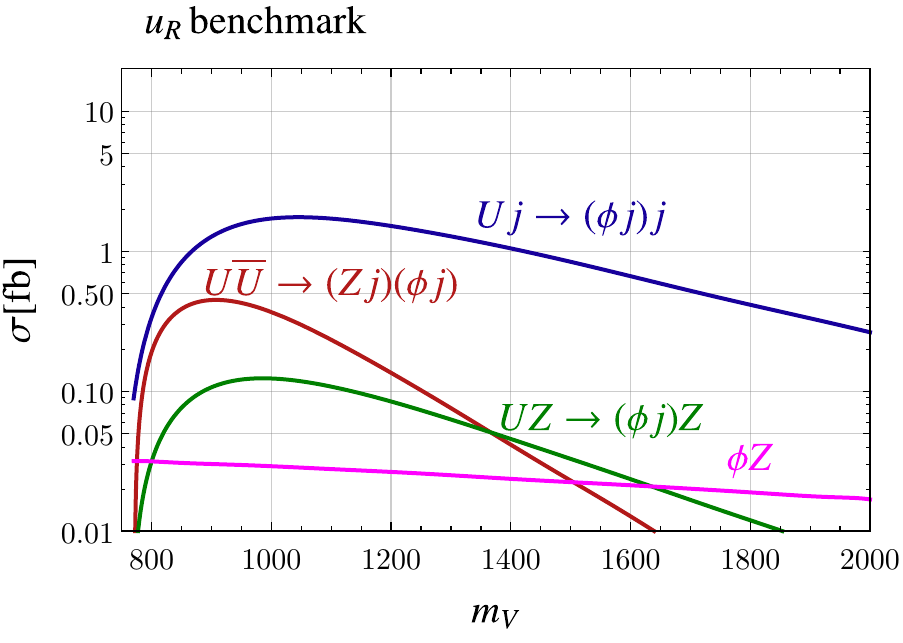}
  \includegraphics[width=7cm]{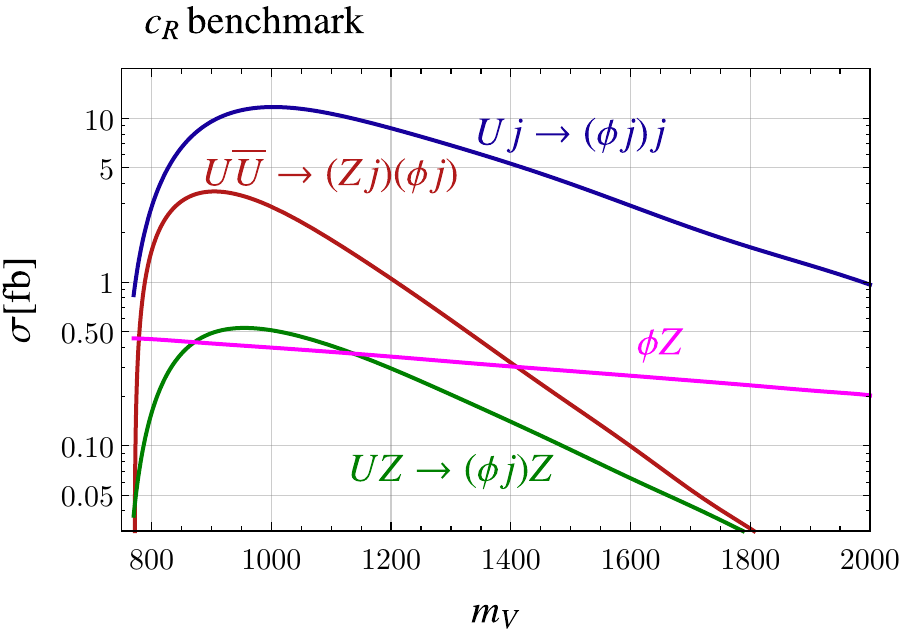}
\caption{The cross sections for the different production modes at our benchmark points. We see that production through single VLQ dominates with the secondary production modes providing up to 10-30\% corrections on the inclusive diphoton cross section.}
\label{fig:comparingXS}
\end{figure*}

In addition to the features of the dominant production mode, there are secondary production modes of the excess. In principle, one may expect that any of the VLQs could decay into the resonance, but this is not the case. Due to the custodial structure of the model, only $ U $ couples to the resonance and a SM quark (see equation~\ref{eq:USu}), and hence its the only single VLQ production mode. However, there are three subdominant modes which can contribute significantly to the cross section, single VLQ production through $p p \rightarrow U Z, \overline{ U} Z $, QCD pair production of VLQs through $ p p \rightarrow U \overline{ U} $, and direct production of the resonance through a $ t $-channel VLQ, $ p p \rightarrow \phi Z $. The cross section composition depends strongly on the choice of mixing angle and mass of the VLQ. The various contributions to the total cross section as a function of the mass for our benchmark points are shown at leading order in figure~\ref{fig:comparingXS} (for simplicity we do not apply $ K $-factors when comparing between these different channels). We conclude that the additional production modes make up 10-30\% of the inclusive diphoton cross section for reasonable choices of parameters. With more statistics, excesses in these subleading channels could be used to differentiate our signal.

Lastly we note that in principle the charged scalars, which are almost degenerate with $ \phi $, can also be observed as they are singly produced by a similar mechanism as $ \phi $. However, the loop-induced decays to $ WZ $ and $ W \gamma $ both vanish in the custodial limit, rendering their $ 3 $-body decays dominant in almost all of parameter space. These $ 3 $-body decays could be probed, but such studies are likely less sensitive then other searches.
\section{Conclusions}
We have presented a model describing a 750 GeV diphoton resonance arising from a custodial triplet which is produced as a decay product of a singly-produced VLQ. Our model has novel kinematics compared to other proposed production mechanisms and eliminates the tension from the 8 TeV diphoton searches while maintaining consistency with the kinematic distributions in the excess region. With additional statistics, our signal could be confirmed by the presence of a forward jet in the diphoton events or as a kinematic edge in the leading-jet $p_T$ distributions if the VLQ mass is significantly heavier than 800 GeV. The scalar resonance enjoys custodial protection, explaining the dominance of the $\gamma\gamma$ decay rate over $WW$, $hh$, and dijet decays.

Additional signatures of the model include a corresponding excess in the $Z \gamma$ and $ Z Z $ channels with fixed rates with respect to the $ \gamma \gamma $ rate. Furthermore, searches for single production of VLQs in Run II will probe deep into the viable parameter space of this model. 
 
We now note some interesting model building possibilities which we leave for future studies. First, in this work we focus on the case where the new scalar arises from an $ \text{SU}(2) _R $ triplet but is uncharged under $ \text{SU}(2) _L $. However, many of the benefits enjoyed by this model are present in similar representation choices, in particular if $ \phi  $ is a $ ( {\mathbf{3}} , {\mathbf{1}}  ) $ or $ ( {\mathbf{3}} , {\mathbf{3}} )$ under $ ( \text{SU}(2) _L , \text{SU}(2) _R ) $. These models also forbid gluon fusion and the tree-level production mechanism can dominate. Another interesting possibility is if the VLQs are related to the the top sector, as one might expect in a composite Higgs model, and the custodial symmetry is broken explicitly by the top Yukawa. Such a breaking can induce $\phi$ production through gluon fusion, perhaps in a controlled manner such that the decays to $hh$ as well as decays to $ W ^+ W ^- $ are still suppressed. Finally, we comment on some ways to further reduce the size of the Yukawas. In this work we focused on a single flavor of VLQ for simplicity. However, if there are additional flavors (with or without mixing to the other SM quarks), this can greatly enhance the diphoton decay rate, reducing the size of Yukawas necessary to reproduce the excess. An additional possibility is if SU(2)$_R $ is gauged. In this case, the additional gauge bosons will propagate in the diphoton loop giving a significant enhancement to the diphoton rate. 

\acknowledgments{We thank Eric Kuflik for collaboration in the early stages of this project. We are grateful to Kiel Howe and Wee Hao Ng for useful discussions. This work was supported in part by the NSF through grant PHY-1316222. JD is supported in part by the NSERC Grant PGSD3-438393-2013.}

\appendix

\section{Model details\label{app:couplings}}
\subsection{Couplings}
In this section, we derive the couplings relevant for the model between the quarks and the vector-like quarks, beginning with the $ Z $ couplings. 
 The $Z$ boson interactions with the up-type quarks in the interaction basis are given by (we define $ {\slashed Z} \equiv \bar{\sigma} ^\mu Z _\mu $)
\footnote{Note that $U_1$ is the upper component of an SU(2)$_L$ doublet, while $U_2$ is the lower component of a second doublet. }
\begin{widetext}
\begin{align} 
{\cal L} _Z & = \frac{ e }{ c _w s _w } \left\{ \frac{2}{3} s _w ^2 \overline{u} ^{(0)\, \dagger } {\slashed Z} \overline{u} ^{(0)}   +\left( \frac{1}{2} + \frac{2}{3} s _w ^2 \right) \overline{U}^ {  \dagger }  _1 {\slashed Z} \overline{U}_1 + \left( - \frac{1}{2} + \frac{2}{3} s _w ^2 \right) \overline{U} _2 ^{ \dagger } {\slashed Z} \overline{U} _2  \right\} \\ 
& = \frac{ e }{ s _w c _w } \mathbfcal{U} ^{(0)\dagger}  \left( \begin{array}{ccc} 
 \frac{2}{3}  s _w ^2  & 0 & 0 \\  
0 & \frac{1}{2} + \frac{2}{3} s _w ^2  & 0 \\  
0 & 0 & -  \frac{1}{2} + \frac{2}{3} s _w ^2    
\end{array} \right) {\slashed Z} \mathbfcal{U}^{(0)} \,,   \quad \mathbfcal{U}^{(0)} \equiv \left( \begin{array}{c} 
\overline{u} ^{(0)}   \\  
\overline{U}_1 \\  
\overline{U}_2   
\end{array} \right) .
\end{align}
\end{widetext}
Notice that we can split the coupling matrix into two pieces,
\begin{equation} 
\frac{2}{3} s _w ^2 \left( \begin{array}{ccc} 
1 & 0 & 0 \\  
0 & 1 & 0 \\  
0 & 0 & 1  
\end{array} \right) + \frac{1}{2}  \left( \begin{array}{ccc} 
0 & 0 & 0 \\  
0 & 1  & 0 \\  
0 & 0 & -1   
\end{array} \right) .
\end{equation} 
The first matrix is diagonal and commutes with the rotation to the mass basis while the second matrix yields new couplings between the VLQs and the quarks upon moving to the mass basis. Performing the rotation (the rotation matrices are given in eq.~\ref{eq:Rmix}) gives,
\begin{align} 
{\cal L} _Z & = \frac{ e }{ c _w s _w }  \bigg\{\frac{1}{2}\mathbfcal{U} ^\dagger \left( \begin{array}{ccc} 
0 & 0 &  - s _\theta  \\  
0 & 0 &  - c _\theta  \\  
 -  s _\theta  &  -  c _\theta  & 0  
                                                          \end{array} \right)  {\slashed Z} \mathbfcal{U} + \frac{2}{3} s  _w ^2 \mathbfcal{U} ^\dagger  {\slashed Z}  \mathbfcal{U} \bigg\}\,, \notag \\ 
 \mathbfcal{U} & \equiv \left( \begin{array}{c}
\overline{u}  \\  
\overline{\tilde{U}}  \\  
\overline{U} 
\end{array} \right).
\end{align} 
Notice that the rotation to the mass basis has left the top-left entry of the coupling matrix unchanged. This is very important as it means the mixing with the VLQs does not affect the $ Z \bar{u} \bar{u} $ coupling which is tightly constrained experimentally. We see that we have a new coupling between the quark and the VLQ:
\begin{equation} 
{\cal L} _{Zu U} =  - \frac{ e }{ c _w s _w } \frac{1}{2} s _\theta  \left( \overline{u} ^\dagger {\slashed Z}   \overline{U} \right) + h.c.
\end{equation} 

Now consider the $ W $-boson couplings. The right-handed up quark does not couple to the $ W $ in the gauge basis, so the relevant couplings are simply:
\begin{align} 
{\cal L} _{W} & =  - \frac{ g }{ \sqrt{2} } \left( \overline{U} ^{ \dagger }_1 {\slashed W} ^-   \overline{D} + \overline{X} ^{ \dagger } {\slashed W} ^-  \overline{U}_2\right)  + h.c.\\ 
& \subset - \frac{ g s _\theta}{ 2 } \left( \overline{u} ^\dagger {\slashed W} ^- \overline{D} +  \overline{X}^\dagger {\slashed W} ^- \overline{u}  \right)  + h.c.
\end{align}
where we moved to the mass basis in the last line.
\subsection{Decay rates\label{app:decays}}

In this section, we present formulae for the different decay rates used in the text for both the scalar resonance and the $ U $ quark.

\subsubsection{$ \phi $ decays}
We begin by considering the tree-level decays of $ \phi $. The dominant contribution is
\begin{equation} 
\Gamma ( \phi \rightarrow Z u \bar{u} )  = \frac{m_\phi N _c }{ 4 ( 4\pi ) ^3  }   \frac{ m _Q ^2  }{ v ^2  } s  _\theta ^4   \,y _\phi ^2 \,  g_Z ( \tau ) 
\end{equation} 
where $ N _c $ is the number of colors, $ \tau \equiv m _Q / m_\phi$, and  
\begin{equation} 
g _Z( \tau )\equiv \int _0 ^1  d x \int  _{ 1 - x } ^1 d \bar{x}   \frac{ ( 1 - x ) ( 1 - \bar{x} )  ( 2 - x - \bar{x} - 2\tau ^2 ) ^2  }{ ( 1 - x - \tau ^2 ) ^2  ( 1 - \bar{x} - \tau ^2 )  ^2 }  .
\end{equation}  
The other conceivable $3$-body decays, $ \phi \rightarrow h u \bar{u} $, $ \phi \rightarrow W d \bar{u} $  , and $ \phi \rightarrow W u \bar{d} $ vanish identically due to the custodial production. 

In addition $ \phi $ has several loop induced decays to vector bosons as well as the Higgs. All the loop induced decays decays violate custodial symmetry. This can easily be seen at the operator level, where the terms 
\begin{equation} 
\Phi B _{ \mu \nu } B ^{ \mu \nu  } , \ \ \Phi W _{ \mu \nu } ^a W ^{ \mu \nu  }  _a , \ \ \text{and} \ 
\ \Phi G _A ^{  \mu \nu } G ^A _{ \mu \nu }
\end{equation} 
(where $ B _{ \mu \nu}, W _{ \mu \nu } ^a , $ and $ G _{ \mu \nu} ^A $ represent the hypercharge, $ \text{SU(2)} _L $, and QCD field strength tensors respectively) all violate custodial symmetry and $ \text{Tr} \left[{\cal H} ^\dagger \Phi      {\cal H} \right]    $ vanishes identically. There is a large breaking of this symmetry from gauging hypercharge, which induces decays into $ \gamma \gamma $, $ \gamma Z $, and $ ZZ $. Since gauging hypercharge only breaks $ SU(2) _R $, the $ Z $ interactions are suppressed by powers of the Weinberg angle, resulting in these being generically subdominant to the photon decays. The general computation of these decay rates is made complicated due to the mixing of the VLQs with the up quark, however since these contributions are suppressed by powers of the mixing angle they are generically small. We have checked the size of these corrections by computing the rates numerically using {\scshape FeynArts3}, {\scshape FormCalc8}, and {\scshape LoopTools2}~\cite{Hahn:1998yk,Hahn:2000kx} and we find that the effect is at most $ 10\% $ in the interesting region of parameter space (though often much smaller), and we neglect these effects for simplicity. 

The decay rate of a scalar into two photons mediated by VLQs with mass $ m _i $ is~\cite{Gunion:1989we}
\begin{equation} 
\Gamma ( \phi \rightarrow \gamma \gamma ) = \frac{ m_\phi^3N _c ^2  }{ 4 ( 4\pi ) ^5 }  e ^4  \Big( \sum _i \frac{ y _\phi ^{i  } }{ m _i  } Q _i ^2 A _{ 1/2} ( x _i   )  \Big) ^2 \, ,
\end{equation} 
where $ x _i \equiv 4 m _i ^2 / m_\phi^2 $ and (for $ m _i > m_\phi/ 2 $) $ A _{ 1/2} ( x ) = 2 x ( 1 + ( 1 - x ) \arcsin(1/\sqrt{ x} ) ^2 )  $. The sum runs over all VLQs and for a bidoublet the sum is
\begin{equation} 
 \sum _i \frac{ y _\phi ^{i  } }{ m _i  }Q _i ^2  A _{ 1/2 } ( x _i ) = \left[ \frac{y _\phi}{ m _V } \frac{ 8 }{ 3} \right] A _{ 1/2} ( x _V  )    \, ,
\end{equation} 
where the only non-zero contribution arises from the $ D $ and $ X $ quarks. 

The decay to two gluons mediated by VLQs is
\begin{equation} 
\Gamma ( \phi \rightarrow g g ) = \frac{ m_\phi^3  }{ 2 ( 4\pi ) ^5 } g _s ^4  \Big( \sum _i \frac{ y _\phi ^{i  } }{ m _i  } A _{ 1/2} ( x _i   )  \Big) ^2 \,,
\end{equation} 
where $ g _s $ is the strong coupling constant. For a bidoublet of VLQs with a triplet scalar, the sum is equal to zero showing that gluon fusion is custodially protected as expected.

The decay to $ ZZ $ mediated by VLQs is
\begin{equation} 
\Gamma ( \phi \rightarrow ZZ ) = \frac{  m_\phi^3N _c ^2  }{ 4 ( 4\pi ) ^5 } \frac{ e ^4  }{ s _w ^4 c _w ^4 }  \Big( \sum _i \frac{ y _\phi ^{i  }  }{ m _i  } A _{ 1/2} ( x _i   )  ( T _3 ^i - Q _i s _w ^2 ) \Big) ^2 \,.
\end{equation} 
The sum for the bidoublet is:
\begin{equation} 
 \sum _i \frac{ y _\phi ^{i  } }{ m _i ^2 } A _{ 1/2} ( x _i   )  ( T _3 ^i - Q _i s _w ^2 ) ^2   = \left[ \frac{ y _\phi  }{ m _V } \frac{ 8  }{ 3}s _w ^4 \right] A  _{ 1/2} ( x _V ).
\end{equation} 

The decay to $ Z \gamma  $ is
\begin{align} 
\Gamma ( \phi \rightarrow Z \gamma  ) & =  \frac{  8 m_\phi^3 N _c ^2   }{  ( 4\pi ) ^5 } \frac{ e ^4 }{ s _w ^2 c _w ^2 } \bigg( \sum _i  \frac{ y _i ( T _3 ^i - Q _i s _w ^2 ) Q _i   }{ m _i } \notag \\ 
& \hspace{1cm} \times \left( I _1 ( x _i , \lambda _i ) - I _2 ( x _i,  \lambda _i ) \right)        \bigg) ^2  \, 
\end{align} 
where $  \lambda _i  \equiv  4 m _i ^2 / m _Z  ^2   $ and
\begin{align} 
& I _1  ( a , b ) \equiv \frac{ a b }{ 2 ( a - b ) } + \frac{ a ^2 b ^2 }{ 2 ( a - b ) ^2 } \left[ f ( a ) ^2 - f ( b  ) ^2 \right] \notag \\ 
& \hspace{2cm} + \frac{ a ^2 b }{ ( a - b ) ^2 } \left[ g ( a ) - g ( b ) \right] , \\ 
& I _2 ( a , b ) \equiv - \frac{ a b }{ 2 ( a - b ) } \left[ f ( a ) ^2 - f ( b ) ^2 \right] ,
\end{align} 
and $  f ( x ) \equiv \sin ^{-1} ( 1 /\sqrt{x} ) $ and $ g ( x ) \equiv \sqrt{ x - 1 } f ( x ) $. For the bidoublet,
\begin{align} 
 & \sum _i \frac{ y _i }{ m _i } Q _i  ( T _3 ^i  - Q ^i  s _w ^2 ) ( I _1  ( x _i , \lambda _i ) - I _2 ( x _i , \lambda _i ) )   =  \notag \\ 
& \quad -\left[  \frac{ y _\phi }{ m _i }  \frac{ 8 }{ 3}s _w ^2 \right] ( I _1  ( x _V , \lambda _V ) - I _2 ( x _V , \lambda _V ) ) .
\end{align} 
The $ \Phi \rightarrow \gamma \gamma , \ Z \gamma , \ ZZ $ decays obey the expected relationship when they all arise from $ \Phi B _{ \mu \nu } B ^{ \mu \nu }$:
\begin{equation} 
1 \, : \, 2 \tan ^2  \theta _w \, : \, \tan ^4 \theta _w 
\end{equation}

The decay of $ \phi $ to $ W ^+ W ^-  $ is
\begin{equation} 
\Gamma ( \phi \rightarrow W ^+ W ^-  ) = \frac{  m_\phi^3N _c ^2  }{ ( 4\pi ) ^5 } \frac{ e ^4 }{ s _w ^4 }   \Big( \sum _i \frac{ y _\phi ^{i  } }{ m _i  } A _{ 1/2} ( x _i   ) \Big) ^2 \,.
\end{equation} 
For a bidoublet the sum vanishes identically as expected.

Lastly, the $ \phi h h   $ operator vanishes at tree-level and at one-loop by custodial symmetry but will be generated at two-loops by custodial symmetry breaking.

\subsubsection{$ U ^- $ decays}
\label{sec:Udecay}
The vector-like quarks can decay in a couple ways. We will assume $ m _{ Q} > m _\phi $ such that the VLQs can decay to the scalar. Furthermore, we will focus on $ U  $ since that's the only VLQ that will play a role in the phenomenology.

\begin{figure}[!t] 
  \begin{center} 
\includegraphics[width=7cm]{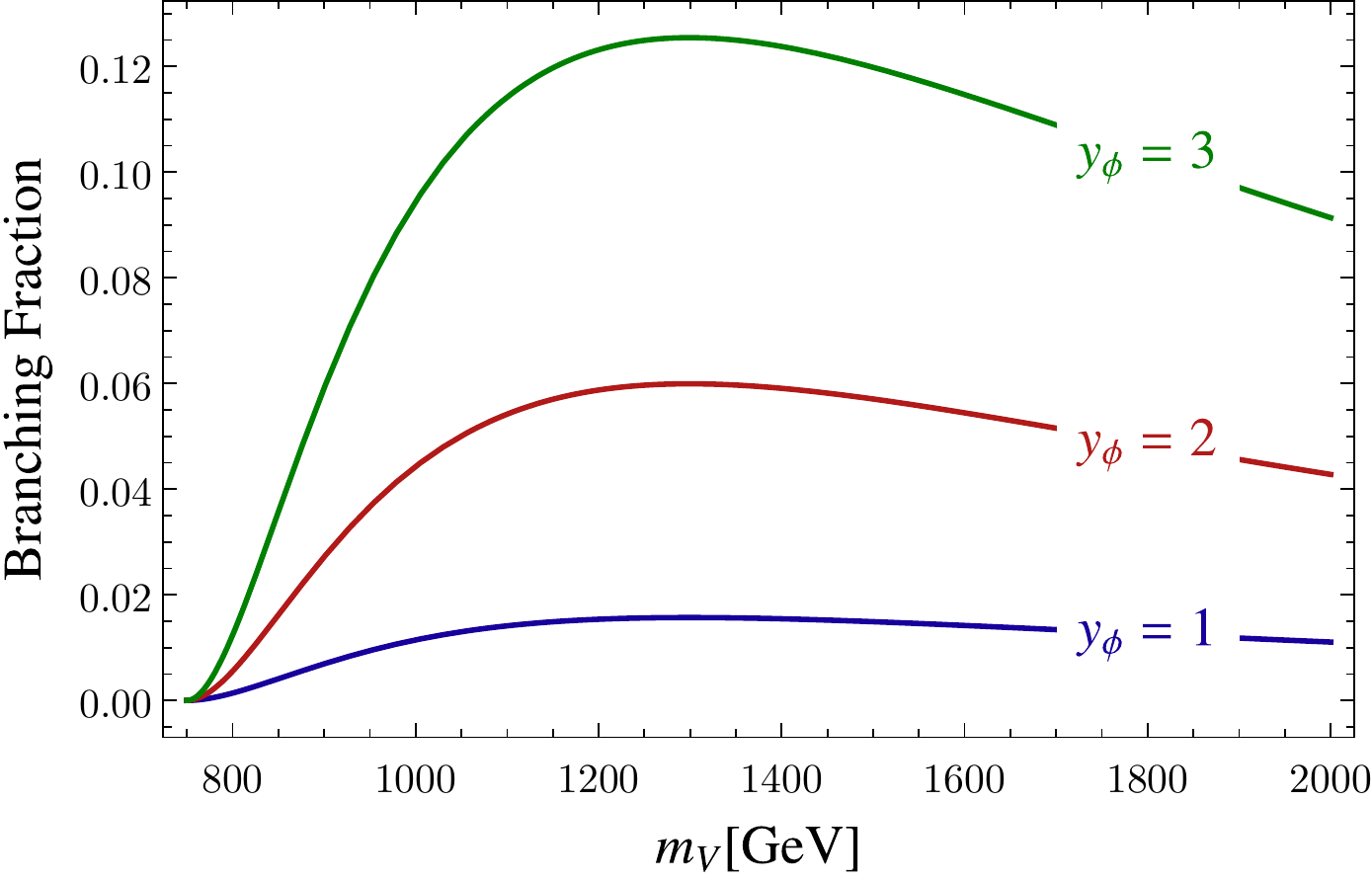} 
\end{center}
\caption{Branching ratio of the $ U \rightarrow u \phi $ decay for different value of the couplings. The fraction is independent of the mixing angle.}
\label{fig:Ubranching}
\end{figure}

The decays rates are
\begin{align} 
& \Gamma (  U  \rightarrow uZ )  \approx   \frac{ m _V }{ 64 \pi }\frac{ e ^2 s _\theta ^2 }{ s _w c _w }  \frac{ m _V ^2 }{ m _Z  ^2 } \\ 
& \Gamma (  U  \rightarrow u \phi )  = \frac{ m _V  }{ 32 \pi }\,y _\phi  ^2 \,s _\theta ^2  \left( 1 -  \frac{ M ^2 _\phi}{ m _V ^2 }  \right)  ^2 
\end{align} 
Notice that the $ u Z $ decay is enhanced by $ m _V ^2 / m _Z ^2 $ due to the longitudinal polarization of the $ Z $. Thus in order for the $ \phi $ decay to be substantial one needs larger Yukawas. The branching ratio into $ \phi\, u $ is shown in figure~\ref{fig:Ubranching} for different Yukawas.

\subsection{Custodial symmetry breaking}\label{app:custodial_breaking}
In this work we have assumed that the couplings and masses of the VLQs and the triplet $\Phi$ preserve the custodial symmetry, which enforces a cancellation in the loop amplitudes corresponding to the gluon fusion production of $\phi$ as well as the decays to $gg$, $hh$, and $WW$. Assuming no cancellations or large mass hierarchies, a generic scalar $\phi $ coupling to $N_f$ VLQs with coupling $y _\phi $ would acquire an effective coupling to gluons of the form
\begin{equation}
\mathcal{L}_{\text{generic}} \supset - \frac{1}{16 \pi^2}\frac{N_f g_s^2 y _\phi }{4 m_\phi} \,\phi\,  G^{\mu \nu, A}G_{\mu \nu}^A \,,
\end{equation}
with similar expressions for the other amplitudes. The explicit breaking of custodial symmetry due to the gauging of hypercharge means that these amplitudes will still be generated, but with an additional suppression of $\mathcal{O}(\alpha/c_w^2)$ compared to the above estimate.

\begin{figure} 
  \begin{center} 
\includegraphics[width=8cm]{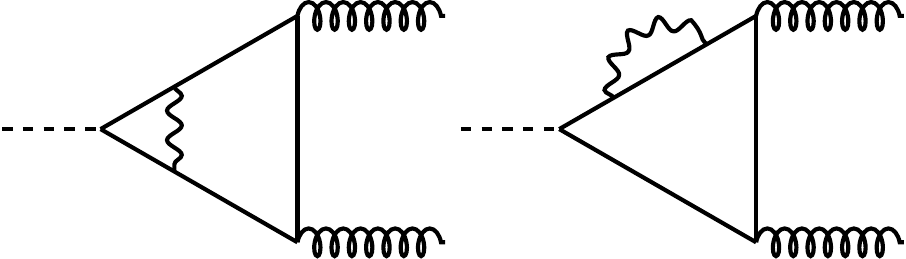} 
\end{center}
\caption{Prototypical loop contributions to the custodial symmetry breaking amplitudes. Such two-loop contributions can induce gluon fusion production and decays two gluons. Similar diagrams can give rise to decays to $ W ^+ W ^- $ and $ h h $.}
\label{fig:custodialloops}
\end{figure}

In particular, the VLQ mass renormalization and the renormalization of the $ \text{Tr} \left[ \overline{V} \Phi V \right] $ couplings due to hypercharge gauge boson loops, illustrated in Figure \ref{fig:custodialloops}, will contribute operators of the form $\text{Tr}[T_3^R \Phi]G^{\mu \nu, A}G_{\mu \nu}^A$. Since the mass and vertex renormalizations are logarithmically divergent, they require counterterms which are not calculable in the effective theory. Instead, we calculate the size of the IR contributions and take this as an estimate of the overall size of the irreducible contributions.

The mass renormalization of the VLQs introduces a mass splitting between the different $T_3^R$ states of size
\begin{align}
\frac{\delta m_V}{m_V} &\simeq \frac{3 g'^2}{16 \pi^2}\log\left(\frac{\Lambda^2}{m_V^2} \right) \Delta \left(Y^2\right)\\
& \simeq \frac{\alpha}{\pi c_w^2}\log\left(\frac{\Lambda^2}{m_V^2} \right).\notag
\end{align}
Similarly, the vertex and wavefunction renormalization provide a contribution to the operator $\delta y _\phi  \, \text{Tr} \left[ \overline{ V} T _{ 3 } ^R \Phi V \right] $ of size 
\begin{align}
\frac{\delta y _\phi }{y _\phi } &\simeq \frac{6 g'^2}{16 \pi^2} \log \left( \frac{\Lambda}{m _V } \right) \Delta \left(Y^2\right)\\
&\simeq \frac{\alpha}{\pi c_w^2} \log \left( \frac{\Lambda^2}{m _V ^2} \right).\notag
\end{align}
where $ g '  $ is the $ U(1) _Y $ coupling constant.
Now, custodial symmetry violating amplitudes of the kind in Figure \ref{fig:custodialloops} can be generated either with an insertion of $\delta m_V$ instead of $m_V$, or with the coupling $\delta y _\phi $. Therefore, the amplitude is suppressed by a factor
\begin{align}
\frac{\delta \mathcal{A}}{\mathcal{A}_0} &\simeq \frac{\delta m_V}{m_V} + \frac{\delta y _\phi }{y _\phi }\\
&\simeq \frac{2 \alpha}{\pi c_w^2}\log\left(\frac{\Lambda^2}{\text{TeV}^2} \right)
\end{align}

In the same spirit, one can also generate a mixing between the Higgs and the new scalar $ \phi $. Such a mixing is induced at two loops from the operator of the form, $ y _{ \Phi H H } m_\phi \text{Tr} \left[ H ^\dagger T _3 ^R \Phi H \right]  $. The coefficient of this operator is of order
\begin{equation} 
y _{\Phi  H H  } \simeq y _\phi \frac{\lambda _V ^2 }{ 16 \pi ^2 } \frac{2 \alpha}{\pi c_w^2}\log\left(\frac{\Lambda^2}{\text{TeV}^2} \right) \,,
\end{equation} 
which results in a mixing angle between the Higgs and $ \phi $ of order
\begin{equation} 
\simeq  \frac{ m _V ^2  \tan ^2 \theta }{  \sqrt{2} v m_\phi } \frac{y _\phi  }{ 16 \pi ^2 } \frac{2 \alpha}{\pi c_w^2}\log\left(\frac{\Lambda^2}{\text{TeV}^2} \right) \,\sim \,{\cal O} ( 10 ^{ -4} - 10 ^{ - 5}) \,,
\end{equation} 
where we have substituted $ \lambda _V $ for the mixing angle. This mixing will induce decays of $ \phi  $ to $ t \bar{t}  $, but due to the smallness of the coupling, we do not expect this decay to be observable in the near future.

\subsection{Down-type model\label{app:down}}
We now present the down-type model which can have mixing between the SM down-type quarks and the VLQs. The model is identical to the up-type model but assigning the $ V $ bidoublet a U(1)$ _X $ charge of $ - 1/3 $ (as opposed to $ + 2/3 $). This gives the following fields  
\begin{equation} 
V = \left( \begin{array}{cc} 
D _2  & U \\  
Y & D _1 
\end{array} \right)  \quad \overline{V} = \left( \begin{array}{cc} 
\overline{D} _2  & \overline{Y}  \\  
\overline{U}  & \overline{D}_1 
\end{array} \right) 
\end{equation}
where $ Q _{ D _1 }  = Q _{ D _2 }= - 1/3 $, $ Q _{ U } = + 2/3 $, and $ Q _Y = - 4/3 $. As in the up-mixing case, a mixing is generated between a SM quark and a VLQ through:
\begin{equation} 
{\cal L} _{ VLQ} = m _V  \text{Tr} \left[ \overline{V} V \right] + \lambda _V \text{Tr} \left[ {\cal H}^\dagger V  \right]  \overline{d} ^{(0)} + h.c.
\end{equation} 
where $ \overline{ d } ^{(0)} $ denotes the down quark in the SM mass basis of the down-type sector. The mixing produces a $ Z D d  $ coupling resulting in electroweak production of $ D $, which can decay into the diphoton resonance. 
\section{Experimental constraints\label{app:constraints}}

\begin{figure} 
\begin{center} 
  \includegraphics[width=8cm]{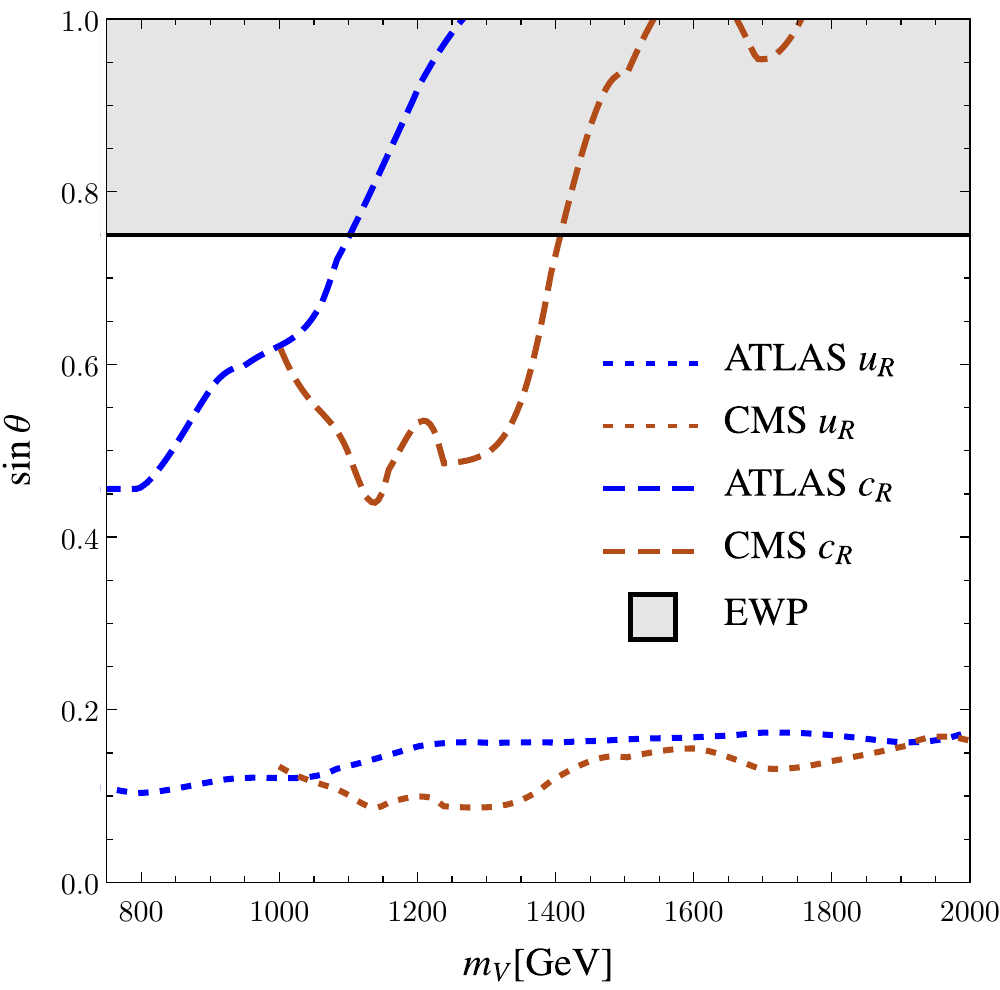}
\end{center}
\caption{The constraints on the bidoublet model (reinterpreted from the work of~\cite{Delaunay:2013pwa}) arising from an ATLAS 7 TeV dedicated search for single production of VLQs~\cite{ATLAS:2012apa}, a CMS 8 TeV search for $W$/$Z$-tagged dijet resonances~\cite{CMS-PAS-EXO-12-024}, and electroweak precision (EWP). Areas above the lines are excluded. Here we neglected effects due to additional decay channels of the vector-like-quarks into the scalars in our model.}
\label{fig:limits}
\end{figure}

The LHC has performed searches with significant sensitivity to models with light-quark mixing. The constraints were studied in detail in~\cite{Delaunay:2013pwa} for both up-quark mixing and charm-quark mixing in the context of a composite model and in~\cite{Atre:2013ap} for the up-type mixing model. The dominant constraints arise from charged current production of $ D $ and $ X $ quarks. There are additional constraints from production of the charged $ + 2/3 $ quarks, but since they are always subdominant, we omit these. Instead of recasting the constraints ourselves we make use of the recast performed in Ref.~\cite{Delaunay:2013pwa}. The authors recast two searches: a 7 TeV search by ATLAS searching for the bidoublet model (without the additional scalar triplet)~\cite{ATLAS:2012apa} and an $ 8 $ TeV search for excited quarks~\cite{CMS:2013fea} with a similar final state which is not optimized for the single production of vector-like-quarks but shares a similar final state. The two searches have competitive limits. Additionally there are constraints on pair production of VLQs, however these are subdominant in the mass ranges we are interested in. In particular ATLAS has performed a search for VLQs decaying to $ W j $ finding a limit around $ 700 \text{ GeV} $ for a single VLQ~\cite{Aad:2015tba}. With two copies of such VLQs the limits strengthen but do not extend past $ 800 \text{ GeV} $. Furthermore, electroweak precision places an additional constraint from additional contributions to the $ S $ parameter~\cite{Atre:2011ae}. One might worry that the additional scalar would complicate the limits, in particular the VLQs can decay to the scalar weakening the constraints. In general these branching ratios are $ \lesssim 10 \% $ and we ignore such effects in our discussion.  

Our goal is now to convert the single production limits quoted Ref.~\cite{Delaunay:2013pwa} into our (closely related) framework. In Ref.~\cite{Delaunay:2013pwa} the authors study a bidoublet model but with an additional VLQ singlet which they denote as $ \tilde{U} $. We can decouple the particle to match with our framework. Multiplying their cross sections by the correction factor,
\begin{equation} 
s _\theta ^2 \left[ \cos \frac{ v }{ f} \sin \left( \tan ^{-1} \left( \frac{ y _R f }{ m _V } \sin \frac{ v }{ f} \right) \right) \right] ^{ - 2}
\end{equation} 
with $ f = 600 \text{ GeV}  $, $ v \simeq 246 \text{ GeV} $, and $ y _R = 1 $ gives the cross sections in our case. Employing this procedure we obtain the limits shown in figure~\ref{fig:limits}. 
\bibliographystyle{JHEP}
\bibliography{Diphoton}{}
\end{document}